\documentstyle[preprint,prb,aps]{revtex}
\tightenlines 
\begin{document}
\preprint{\vbox{\noindent
\null\hfill INFNCA-TH0007}}
\draft
\title{Third order renormalization group applied to
the attractive one--dimensional Fermi gas}

\author{Paolo Carta\cite{email}}
\address{Dipartimento di Fisica Universit\`a di Cagliari,\\
Cittadella Universitaria 09042,
Monserrato, Italy\\
and INFN, sezione di Cagliari.}


\maketitle

\begin{abstract}
We consider a Callan--Symanzik and a Wilson renormalization group (RG)
approach to the infrared problem for interacting fermions in one
dimension with backscattering. We compute the third order (two--loop)
approximation of the beta function using both methods and compare it
with the well known multiplicative Gell--Mann Low approach. We point
out a previously unnoticed strong instability of the third order fixed
point with respect to an arbitrary dimensionless parameter, which
suggests a RG flow toward a strong coupling phase.

\end{abstract}

\pacs{{\tt 71.10.Pm, 71.10.Hf}}


\narrowtext

\section{Introduction}
The problem of the one--dimensional Fermi gas model of a metallic
conductor, in the low energy approximation, has been approached using
three methods: conventional many--body techniques\cite{dya} and,
mainly, the bosonization\cite{schotte,le,lp} and the renormalization
group (RG) methods\cite{mey,sol,sol2,bg,bg2,bgps}. In this paper we
will be concerned with the latter approach.  A formulation of the
Gell--Mann Low multiplicative RG for this problem was introduced in
Ref.\onlinecite{mey}. The model considered was the {\it g--ological
model}, which describes a weakly interacting one--dimensional fermion
system with Tomonaga--type ($g_2$, $g_4$) and backscattering ($g_1$)
interactions. Phonons are neglected.  That method provided a
satisfactory understanding of the infrared behavior in the case of a
weak {\it repulsive} (effective) interaction. A short list of the most
relevant results in this case may be the following (for extensive
reviews see e.g. Refs.\onlinecite{sol,schul}): i) the RG flows toward
the {\it Luttinger liquid}\cite{hal} fixed point\cite{lut,ml}; ii)
there is a line of nontrivial fixed points; iii) in the infrared limit
the system is not asymptotically free, as in the {\it Fermi liquid}
case, but is described by anomalous exponents. These results were
recovered and rigorously proved also in the case of periodic potential
using the Wilson RG\cite{bm1}.

Things change considerably if we consider a weak \textit{attractive}
interaction. Since in this case there is not a second order (one loop)
finite fixed point, in Ref.\onlinecite{sol} the computation of the
beta function was carried to third order (two loops). It was found a
$O(1)$ third--order fixed point. This result, if reliable, would be of
extreme physical interest because it would signal a behavior
completely different from the Luttinger liquid paradigm. One should
expect the opening of a gap in the dispersion relations, while
Luttinger spectrum is gapless, and an exponential decay of the
correlation functions, while in the Luttinger case there is only a
power low decay with increasing distance. The problem is, of course, how
seriously one should take the very existence of an attractive
perturbative fixed point on the basis of the third order result. The
computation of the the fourth order (three--loop) approximation of the
beta function was discussed in Refs.\onlinecite{ting,q1,q2}. A smaller
but still $O(1)$ fixed point was found.  Moreover only the first two
terms of the beta function are universal. The computation of the third
term is useful provided there is some evidence of a perturbatively
tractable phase  interacting attractively. In this case a precise
determination of the renormalized couplings would be important to
compute the response functions.

It it useful to make a comparison with the results obtained with the
bosonization method. With {\it bosonization} it is meant the bosonic
representation of fermion field operators\cite{mat1,mat2,lp,le}. This
method is in some sense the inverse of the one used to solve exactly
the Luttinger model\cite{ml}, where bosonic degrees of freedom are
expressed in terms of fermionic operators. Probably the most important
result of the bosonization is the exact solution of the model with
backscattering\cite{le} ($g_1$ and $g_2$ terms, see below) in the
particular case where $g_1=-\frac{6}{5}\pi$. Actually the decoupling between
charge and spin degrees of freedom, crucial for the exact solution, is
open to questions \cite{hal2}. Moreover there are problems in the
limiting procedure employed \cite{theu} and the ladder operators
restoring the correct occupation numbers\cite{hal} are not
discussed. (A version of the bosonization free from this problems has
been proposed\cite{sei}. It should be noted that this version does not
deal with the crucial backscattering interaction term: in
Ref.\onlinecite{sei} only the Luttinger model is considered). Anyway,
taking for granted the Luther--Emery solution\cite{le}, the RG method
should fill the missing information for value of $g_1$ in the
neighborhood of the exact solution. From the bosonized representation
of the interaction it is not difficult to derive the third order
scaling equations\cite{le,chui} and the response functions calculated
in Ref.\onlinecite{ting} are in good agreement with the results of
Ref.\onlinecite{le}.

From these considerations one may be tempted to give a heuristic
meaning to the large but finite fixed point. In this paper we want to
show that this is not the case. The main point is that even the sign
of the third order fixed point depends on small variations of a
parameter $\gamma$ whose value can be arbitrarily chosen, provided
$\gamma>1$. We will show this both using the Gell--Mann Low (GML) and
the Wilson RG.

The paper is organized as follows. In section II we briefly review the
multiplicative GML approach. We explain why it is useful to check the
results of this approach using other methods. Recasting the
multiplicative procedure into discrete steps, instead of considering
the usual Lie equation, we reach our main conclusion.  In section III
we formulate a Callan--Symanzik (CS) approach to the problem and
compute the beta function in two--loop approximation. The same
computation is proposed in section IV employing the Wilson RG in the
multiscale formulation\cite{gal,bg}.  Finally in section V we come to
the conclusions.

\section{The Gell--Mann Low approach}
We briefly recall the GML multiplicative RG for one dimensional
interacting fermions. We will follow closely
Refs.\onlinecite{mey,sol}, with the only difference that we find
it convenient to adopt a Euclidean formalism.  We consider the
g--ological model, defined as follows.  The kinetic term is taken
linear around the Fermi surface defined by the two points $-k_F$ and
$k_F$:
$$
H_0 = \sum_{k,\omega,\sigma} (\omega
k-k_F)\psi^{+}_{k,\omega,\sigma}\psi^{-}_{k,\omega,\sigma},
$$
where $\psi^{\pm}_{k,\omega,\sigma}$ are creation and annihilation
operators for right moving $(\omega=1)$ and left moving $(\omega=-1)$
fermions with momentum $k$ and spin $\sigma$ ($\sigma=\pm1/2$).  We
choose units such that $v_F$=1 ($v_F$ is the velocity at the Fermi
surface).  The ultraviolet (u.v.) stability is imposed by bandwidth
cutoffs: the momenta are restricted to the intervals $(\omega
k_F-k_{\text{uv}},\omega k_F+k_{\text{uv}})$ for
$\psi^{\pm}_{k,\omega,\sigma}$.  We define $E_0=2k_{\text{uv}}$.  The
interaction Hamiltonian is \widetext
\begin{eqnarray}\label{e:gol}
H_{\text{int}}= &&
\frac{1}{2L}\sum_{k,p,\omega,\sigma,\sigma'}(g_{1\parallel}\delta_{\sigma,\sigma'}+
g_{1\perp}\delta_{\sigma,-\sigma'})\psi^+_{k_1,\omega,\sigma}\psi^+_{k_2,-\omega,\sigma'}\psi^-_{k_2
+2k_F+p,\omega,\sigma'}\psi^-_{k_1,- 2k_F-p,-\omega,\sigma}\nonumber
\\ +&&
\frac{1}{2L}\sum_{k,p,\omega,\sigma,\sigma'}(g_{2\parallel}\delta_{\sigma,\sigma'}+
g_{2\perp}\delta_{\sigma,-\sigma'})\psi^+_{k_1,\omega,\sigma}\psi^+_{k_2,-\omega,\sigma'}\psi^-_{k_2+p,-\omega,\sigma'}\psi^-_{k_1
-p,\omega,\sigma}\nonumber \\ +&&
\frac{1}{2L}\sum_{k,p,\omega,\sigma,\sigma'}(g_{4\parallel}\delta_{\sigma,\sigma'}+
g_{4\perp}\delta_{\sigma,-\sigma'})(\psi^+_{k_1,\omega,\sigma}\psi^+_{k_2,\omega,\sigma'}\psi^-_{k_2+p,\omega,\sigma'}\psi^-_{k_1
-p,\omega,\sigma}.
\end{eqnarray}
\narrowtext
L is the length of the line. The umklapp interaction term ($g_3$) is
neglected since it is important only in the half--filled band case,
which will be excluded. Since $g_{1\parallel} = -g_{2\parallel}$ it is
always possible to take $g_{2\perp}=g_{2\parallel}= g_2$, reducing the
independent couplings to $g_{1\parallel}$, $g_{1\perp}$, $g_2$.  For
the sake of simplicity it is possible to neglect, at least as a first
approximation, $g_4$: we know from the Mattis model\cite{mat3} that
$g_4$ does not change the essence of the problem.

In the Euclidean formalism the free propagator in momentum space is given
by
\begin{equation}\label{e:props}
G_\omega(k)=\frac{1}{-ik_0 + \omega k_1},
\end{equation}
where $k_0$ is the energy, $k_1$ the momentum (measured from the Fermi
surface), $k=(k_0,k_1)$ and $\omega=1\, (-1)$
for right (left) moving fermions.  The renormalization procedure is a
prescription that defines new couplings for a theory with a lowered
u.v.  cutoff $E_0$. In the limit $E_0\to 0$ we obtain the renormalized
couplings. If $G^R_\omega$ is the interacting propagator, the $d$
function is defined by the relation
$$
G^R_\omega(k)=d\left(\frac{k_1}{k_{\text{uv}}}, \frac{k_0}{E_0} \right)
G_\omega(k).
$$
The multiplicative constants $z$ and $z_i$ ($i=1\parallel, 1\perp, 2$)
that relate $d$ and the adimensional vertex functions
$\widetilde\Gamma_i$ for different values of the cutoff are definite
by:
\begin{eqnarray}\label{e:molti}
&&d\left(\frac{k_1}{k_{\text{uv}}^{\prime}},\frac{k_0}{E_{0}^{\prime}},g'
\right) = z
\left(\frac{E_{0}^{\prime}}{E_0},g
\right)d\left(\frac{k_1}{k_{\text{uv}}},\frac{k_0}{E_0},g \right)\nonumber \\
&&\widetilde\Gamma_i\left(\frac{k_j}{k_{\text{uv}}^{\prime}},\frac{w_j}{E_{0}^{\prime}}
,g'\right) =
z_i^{-1}
\left(\frac{E_{0}^{\prime}}{E_0},g \right)
\widetilde\Gamma_i\left(\frac{k_j}{k_{\text{uv}}},\frac{w_j}{E_0}, g
\right) \nonumber \\ && g_{i}^{\prime} =
g_iz^{-2} \left(\frac{E_{0}^{\prime}}{E_0},
g \right)z_i\left(\frac{E_{0}^{\prime}}{E_0},g \right),
\end{eqnarray}
where $E_{0}^{\prime}<E_0$ is the lowered cutoff, $g$ and $g^\prime$
denote respectively the old and the new couplings.  The invariant
couplings $g_i^{R}$ are defined by:
$$
g_{i}^{R}\left(\frac{E}{E_{0}},g \right) = g_i z^{-2}
\left(\frac{E}{E_{0}},g \right)z_i\left(\frac{E}{E_0},g\right).
$$
The $g_i^{R}$ are invariant in the sense that
\begin{equation}\label{e:inv}
g_{i}^{R}\left(\frac{E}{E_{0}^{\prime}},g^\prime\right)=g_{i}^{R}\left(\frac{E}{E_{0}},g\right).
\end{equation}
The couplings $g'_i$ for the theory with u.v. cutoff $E_0^\prime$ are
defined by
\begin{equation}\label{e:neocost}
g_i^\prime = g_{i}^{R}\left(\frac{E_0^\prime}{E_{0}},g\right).
\end{equation}
A differential equation for $g_i^R$ is readily derived and is the
standard Lie equation:
\begin{displaymath}\frac{d}{dx}g_i^R (x,g) = \frac{1}{x}\frac{d}{d\xi}g_i^R\left
(\xi,g^R(x,g)\right)_{\xi=1}
\end{displaymath}
where $x ={E_0^\prime/E_0}$. We are interested in the scaling limit
$x\to 0. $The two--loop result is
\begin{eqnarray}\label{e:lie}
\frac{dg_{1\parallel}^{R}}{dx} &&= \frac{1}{x}\left[ \frac{1}{\pi} 
g_{1\perp}^{R\,2} +
\frac{1}{2\pi^2}g_{1\parallel}^R g_{1\perp}^{R\,2}\right]\nonumber \\
\frac{dg_{1\perp}^R}{dx} &&= \frac{1}{x}\left[ \frac{1}{\pi}
g_{1\parallel}^R g_{1\perp}^R +
\frac{1}{4\pi^2}(g_{1\parallel}^{R2}g_{1\perp}^R
+g_{1\perp}^{R\,3})\right] \nonumber \\
\frac{dg_{2}^R}{dx} &&= \frac{1}{x}\left[ \frac{1}{2\pi}
g_{1\perp}^{R\,2} +
\frac{1}{4\pi^2}g_{1\parallel}^R g_{1\perp}^{R\,2}
\right]. 
\end{eqnarray}
For spin independent interaction ($g_{1\parallel}=g_{1\perp} =g_1$)
the nontrivial fixed point is found for $g_1^\star=-2\pi$. 

We now want to recover this result iterating by discrete steps the
procedure that defines the new couplings when the cutoff is lowered:
we aim to study the dependence on the scaling parameter.
Let $\gamma >1$. In proper units we put $E_0 = \gamma ^0$ and
$g_{i,0}=g_{i}(E_0)$ for $i=1\parallel, 1\perp, 2$. $g_{i,-1}$ is
defined as (see Eq. (\ref{e:neocost}))
$$
g_{i,-1} = g_i^R \left
(\frac{\gamma^{-1}}{\gamma^0},g_{j,0}\right)= g_i^R \left
(\frac{\gamma^{-1}}{E_0},g\right)
$$
where $g_{(0)}=g(E_0)=g$.  The procedure is iterated in the following way:
for $n<0$ we define
$$
g_{i,n-1} = g_i^R \left (\frac{\gamma^{n-1}}{\gamma^n},g_{j,n}\right)
\qquad i=1\parallel, 1\perp, 2.
$$
From Eqs. (\ref{e:inv}) e (\ref{e:neocost}) we have that the 
$g_{i,n}$ for $n=-1,-2 \ldots$ are the couplings corresponding to the
cutoff sequence $\{\gamma^n\}$:
$$
g_{i}^{R}\left(\frac{\gamma^{n-1}}{\gamma^{n}},g_{j,n}\right)=g_{i}^{R}\left(\frac{\gamma^{n-1}}{\gamma^0},g_{j,0}\right)=g_{i}(\gamma^{n-1}). 
$$
In the limit $n \to -\infty$ we get the renormalized couplings. We have:
\begin{eqnarray}\label{e:gamma}
&&g_{1\parallel,n-1}= g_{1\parallel,n}-\frac{g_{1\perp,n}^2}{\pi}\ln
\gamma \nonumber \\&& \qquad \qquad + \frac{1}{\pi^2}
g_{1\parallel,n}g_{1\perp,n}^2\left(\ln^2 \gamma -\frac{1}{2}\ln
\gamma \right) \nonumber \\ &&g_{1\perp,n-1}=
g_{1\perp,n}-\frac{1}{\pi}g_{1\perp,n}g_{1\parallel,n} \ln \gamma
\nonumber \\ && \qquad \qquad
+\frac{1}{2\pi^2}\left(g_{1\parallel,n}^2g_{1\perp,n} + g_{1\perp,n}^3
\right)\left(\ln^2 \gamma -\frac{1}{2}\ln \gamma \right) \nonumber\\
&&g_{2,n-1}= g_{2,n}-\frac{1}{2\pi}g_{1\perp,n}^2 \ln \gamma \nonumber
\\
&& \qquad \qquad 
+\frac{1}{2\pi^2} g_{1\parallel,n}g_{1\perp,n}^2 \left(\ln^2 \gamma
-\frac{1}{2}\ln \gamma \right).
\end{eqnarray}
It is easily checked that in the limit $\gamma \to 1^+$
Eqs. (\ref{e:lie}) are recovered. In general the fixed point depends
on $\gamma$ (let's remember that it is a third order fixed point). For
$\gamma \ne \sqrt e$, if $g_{1\parallel}=g_{1\perp}=g_1$, we have:
\begin{equation}\label{e:fisso_g}
g_1^\star = \frac{\pi}{(\ln \gamma -1/2)}.
\end{equation}
When $\gamma \to 1$ we obtain the previous result
$g_1^\star=-2\pi$. The Lie equations (\ref{e:lie}) should not be
fundamental and we find no reason to use the continuous RG instead of
its discrete version.  The dependence on $\gamma$ will be discussed in
section V.

As a final comment on this method we note that Eqs. (\ref{e:molti})
rely on neglecting small contributes that would not allow one to set
multiplicative relations where the $z$ factors do not depend on the
external momenta. For example in the case of the one--loop
approximation of the four--point vertex function, proportional to
$$
-\frac{1}{2\pi}\ln \left(\frac{k_0}{E_0}\right) +\frac{1}{4\pi}\ln
\left(1+\frac{k_0^2}{E_0^2} \right),
$$
($k_0$ is the external energy, see figure \ref{fig1}), the second term is
neglected. That is to say that the vertex functions can be divided in
scaling and not scaling terms. The first ones are taken into account while
the second are not discussed in \onlinecite{mey,sol}. For this reason we
find useful to check Eqs. (\ref{e:lie}) using other methods.

\section{The Callan--Symanzik approach}
Within the framework of the multiplicative RG, it is not difficult to
formulate a Callan--Symanzik approach for our problem.  We follow
a common procedure: first we renormalize the theory in the u.v. with a
fixed (renormalized ) i.r. cutoff $m$, then we will compute the beta
function and study the i.r. behavior for $m\to0$. This approach,
devised for a Field Theory, in our case may be considered
unnecessary. Nevertheless we consider it a way to support the GML
result.

The i.r. regularized free propagator is defined by inserting a bare
mass $m_0$ in the propagator (\ref{e:props}):
$$
G_\omega(k,m_0^2)=\frac{ik_0+\omega k_1}{k^2 +m_0^2},
$$
where $k^2=k_0^2+k_1^2$ and again $\omega=1\, (-1)$ for right (left)
moving fermions. We know that the Luttinger model with a local
interaction is not renormalizable in the u.v. \cite{gs} (this is also
seen from the exact solution \cite{ml}). In order to impose the u.v.
stability we choose a nonlocal interaction whose strength decreases
with increasing distance.  The interaction Hamiltonian of the model is
\begin{eqnarray*}
&&H_{\text{int}}=\nonumber \\ 
&&\sum_{\omega,\sigma=\sigma'} \int
d^2x\, d^2y\,\psi^+_{x,\omega,\sigma}\psi^+_{y,-\omega,\sigma'}
V_{1\parallel}(x-y) \psi^-_{y,\omega,\sigma'}\psi^-_{x,-\omega,\sigma}
\nonumber \\ 
&&\sum_{\omega,\sigma \ne \sigma'} \int d^2x\,
d^2y\,\psi^+_{x,\omega,\sigma}\psi^+_{y,-\omega,\sigma'} V_{1\perp}(x-y)
\psi^-_{y,\omega,\sigma'}\psi^-_{x,-\omega,\sigma} \nonumber \\
+&&\sum_{\omega,\sigma , \sigma'} \int
d^2x\,d^2y\,\psi^+_{x,\omega,\sigma}\psi^+_{y,-\omega,\sigma'}
V_2(x-y) \psi^-_{y,-\omega,\sigma'}\psi^-_{x,\omega,\sigma}
\nonumber\\ +&& \sum_{\omega,\sigma \ne \sigma'} \int
d^2x\,d^2y\,\psi^+_{x,\omega,\sigma}\psi^+_{y,\omega,\sigma'} V_4(x-y)
\psi^-_{y,\omega,\sigma'}\psi^-_{x,\omega,\sigma},
\end{eqnarray*}
where $\psi^{\pm}_{x,\omega,\sigma}$ are the fermion field operators
in coordinate space. The potentials $V_i$ may be chosen for instance as
follows:
\begin{equation}\label{e:def_p}
V_i(x)= \frac{g_i}{4}\,p\,e^{-p|x_1|}\delta(x_0),\quad i=1\parallel,1\perp,2,4,
\end{equation}
where $p>0$ is fixed; $x_0$ and $x_1$ are the time and space
coordinates. In momentum space the model is the same as (\ref{e:gol})
with the only difference that the bandwidth cutoffs are replaced by
the nonlocal couplings
$$
g_i \to g_i\,\frac{p^2}{k_1^2+p^2},
$$
where $k_1$ is the exchanged momentum in the given interaction vertex
and $p$ is introduced in Eq. (\ref{e:def_p}).
In the limit $p\to \infty$ we recover the local couplings
of Eq. (\ref{e:gol}).

Fermion loops are logarithmically divergent. The theory is regularized
introducing a cutoff $\Lambda$ by the means of the standard Schwinger
parametrization:
$$
\frac{1}{k^2+m_0^2}=\int_{0}^\infty d\alpha\,
e^{-\alpha(k^2+m_0^2)}\rightarrow \int_{\Lambda^{-2}}^\infty d\alpha\,
e^{-\alpha(k^2+m_0^2)}.
$$
In order to renormalize the theory we find it convenient to follow the
scheme for the local ($p=\infty$) case, even if when $p$ is finite we
make more subtractions than strictly necessary. It is a simple
exercise of standard power counting to find the superficial degree of
divergence $D$ for the $n$--point vertex functions in the local case:
$$
D(\Gamma_n)=2-\frac{n}{2}. 
$$
We renormalize the couplings ($g_i \to g_i^R$), the mass ($m_0\to m$)
and the wave function ($\psi^\pm \to \psi^{\pm R}$). The
multiplicative constant $Z$ is formally introduced by the relation
$\psi^\pm = Z^{1/2}\psi^{\pm R}$. Let $\Gamma^R_{n}$ be the
renormalized proper $n$--point vertex functions. The relation between
bare and renormalized vertex functions is:
\begin{equation}\label{e:Zcs}
\Gamma_{n}({\mathbf
q},m_0,g,\Lambda)=Z^{-n/2}\Gamma_{n}^R({\mathbf q},m,g^R),
\end{equation}
where $\mathbf{q}$ denotes the $n-1$ independent external momenta of
$\Gamma_n$ and $g =\{g_{1\parallel},g_{1\perp},g_2, g_4\}$.  Of course
the $\Gamma_n$ are functions of the spin and $\omega$ indices attached
to the external fields. For simplicity we have not indicated this
explicitly in Eq. (\ref{e:Zcs}). For the four--point functions the
different possible cases are labeled by a single index
$i=1\parallel,1\perp,2,4$.  It should be noted that while
$\Gamma_{4,i}$ do not depend on the value of $\omega$, $\Gamma_2$
does, so we need an $\omega$ label for this vertex function.  It
proves useful to introduce the reduced two--point vertex function
$\widehat\Gamma_{2}(k)$:
\begin{equation}\label{e:red}
\widehat\Gamma_{2}(k) = (ik_0+\omega k_1)\Gamma_{2,\omega}(k). 
\end{equation}
In the local case $\widehat\Gamma_{2}(k)$ does not depend on
$\omega$. In the non local case the non vanishing terms in the infrared
limit will be $\omega$ independent, so we will neglect the $\omega$
dependence of $\widehat\Gamma_{2}(k)$.  The normalization conditions,
which define $g_i^R$, $m$ and the finite part (zero--loop term)
of $Z$, are:
\begin{mathletters}\label{e:norm}
\begin{eqnarray}
\Gamma_{4,i}^R(\mathbf{0})&&=g_i^R\\
\label{e:norm_2}\widehat\Gamma_2^R(0)&&=m^2\\
\frac{1}{2k_0}\frac{\partial}{\partial
k_0}\widehat\Gamma_2^R(k)\Bigr\vert_{k=(0,0)}&&=1.
\end{eqnarray}
\end{mathletters}
The CS equations are derived considering insertions of operators
related to the derivatives of the vertex functions respect to the
i.r. cutoff $m_0$. To this end we introduce the operator $O$:
$$
O(z) = \sum_{\omega,\sigma}\int \frac{d^2x}{2\pi}\frac{\psi_{x,\omega,\sigma}^+
\psi^-_{z,\omega,\sigma}}{x_0-z_0-i\omega(x_1-z_1)}. 
$$
The corresponding source term in the action has the form $\int d^2x \,
v(x)O(x)$ with $[v]=2$. In momentum space the operator $O$ is
$$
\widetilde O(q)=
\sum_{\omega,\sigma}\int\frac{d^2k}{(2\pi)^2}\frac{\psi_{k+q,\omega,\sigma}^+\psi^-_{k,\omega,\sigma}}{[i(k_0+q_0)+\omega(k_1+q_1)]},
$$
where $\psi_{k,\omega,\sigma}$ are the field operators in momentum space,
and $q$ is the external momentum of the inserted $\widetilde O$
operators. When $O$ is inserted in a vertex function
$\Gamma_n$, the value of $D$ for $\Gamma_{n,O}$ is
$$
D(\Gamma_{n,O})=2-\frac{n}{2}+([O]-2)=-\frac{n}{2}.
$$
The previous relation means that no new u.v. divergences appear due to
$O$ insertions (we do not consider $\Gamma_{0,O}$. We remember that
the vertex functions with insertions are defined as usual by the
Legendre transformation on the field source only and not on the source
of the inserted operators). Since $O$ does not introduce new
divergences, we have $O=ZO^R$. The insertion of $s$ operator $O$ in a
vertex function $\Gamma_n$ will be denoted with $\Gamma_{n,s}$.  In
analogy with Eq. (\ref{e:red}) we define
$\widehat\Gamma_{2,s}(q,{\mathbf k})$ and
$\widehat\Gamma^R_{2,s}(q,{\mathbf k})$, where ${\mathbf k}$ denotes
the $s$ external momenta of the $s$ inserted $\widetilde O$ operators.
Equation (\ref{e:Zcs}) generalizes into
\begin{equation}\label{e:Zcs_gen}
\Gamma_{n,s}({\mathbf
q},{\mathbf k},m_0,g,\Lambda)=Z^{-n/2}Z^{s}\Gamma_{n,s}^R({\mathbf
q},{\mathbf k},m,g^R).
\end{equation}
It is easily deduced that
\begin{equation}\label{e:derCS}
\frac{\partial}{\partial
m_0^2}\Gamma_n({\mathbf q})=\Gamma_{n,1}({\mathbf q},0),
\end{equation}
where the insertion of $\widetilde O$ in the r.h.s. is made at zero
momentum, as indicated. From (\ref{e:Zcs_gen}), (\ref{e:derCS}) and
(\ref{e:red}) we have 
\begin{eqnarray}\label{e:cond_2}
&&\widehat\Gamma_{2,1}^R(0)=1\\
&&m^2=Z m_0^2 \nonumber,
\end{eqnarray}
where the second relation follows from
$\widehat\Gamma^R_{2}(0)=Z\widehat\Gamma_{2}(0)$.  From
Eq. (\ref{e:derCS}) we have:
\begin{equation}\label{e:passo1}
m\frac{\partial}{\partial m}\Gamma_n({\mathbf
q},m_0,g,\Lambda)\biggl\vert_{g,\Lambda}= m\frac{\partial
m_0^2}{\partial m}\biggl\vert_{g,\Lambda}\Gamma_{n,1}({\mathbf
q},0,m_0,g,\Lambda).
\end{equation}
From equations
(\ref{e:passo1}) and (\ref{e:Zcs_gen}), with the
definitions
\begin{eqnarray*}
\gamma_1 &&= \frac{1}{Z}m\frac{\partial Z}{\partial
m}\biggl\vert_{g,\Lambda} \\
\beta_i &&= m\frac{\partial g_i^R}{\partial m}\biggl\vert_{g,\Lambda}, 
\end{eqnarray*}
($\gamma_1$, defined in the previous equations should not be confused
with the RG rescaling factor $\gamma$) we obtain:
\begin{eqnarray}\label{e:CS-1}
&&\left(m\frac{\partial}{\partial m} + \sum_i\beta_i(g^R)\frac{\partial}{\partial
g_i^R} -\frac{n}{2}\gamma_1\right)\Gamma_n^R({\mathbf
q},m,g^R)=\nonumber \\
&&\qquad \qquad Z\left(m\frac{\partial m_0^2}{\partial
m}\biggl\vert_{g_0,\Lambda}\right)\Gamma_{n,1}^R({\mathbf
q},0,m,g^R).
\end{eqnarray}
It is easy to eliminate any reference to the bare theory.  From
Eq. (\ref{e:CS-1}) written for $n=2$ we have:
\begin{eqnarray*}
&& \left(m\frac{\partial}{\partial m} +
\sum_i\beta_i(g^R)\frac{\partial}{\partial g_i^R}
-\gamma_1\right)\widehat\Gamma_2^R({q},m,g^R)\\&& \qquad
\qquad Z\left(m\frac{\partial m_0^2}{\partial
m}\biggl\vert_{g_0,\Lambda}\right)\widehat\Gamma_{2,1}^R({
q},0,m,g^R).
\end{eqnarray*}
From Eqs. (\ref{e:norm_2}) and the first of (\ref{e:cond_2}) we
conclude
$$
Z\left(m\frac{\partial m_0^2}{\partial
m}\biggl\vert_{g_0,\Lambda}\right) = (2-\gamma_1)m^2,
$$
 so that
Eq. (\ref{e:CS-1}) can be written as
\begin{eqnarray*}
&&\left(m\frac{\partial}{\partial m} +
\sum_i\beta_i(g^R)\frac{\partial}{\partial g_i^R}
-\frac{n}{2}\gamma_1\right)\Gamma_n^R({\mathbf q},m,g^R)=\\&&\qquad \qquad
\left(2-\gamma_1\right)m^2 \Gamma_{n,1}^R({\mathbf q},0,m,g^R).
\end{eqnarray*}
The generalization of Eq. (\ref{e:CS-1}) to the case of $s$ insertions
is immediate:
\begin{eqnarray*}
&&\left(m\frac{\partial}{\partial m}\! +\!
\sum_i\beta_i(g^R)\frac{\partial}{\partial g_i^R}
+\left(-\frac{n}{2}+s\right)\gamma_1\right)\Gamma_{n,s}^R({\mathbf
q},{\mathbf k},m,g^R)\\
&&\qquad \qquad = \left(2-\gamma_1\right)m^2
\Gamma_{n,s+1}^R({\mathbf q},{\mathbf k},0,m,g^R).
\end{eqnarray*}
Having set the general definitions and relations of the CS approach we
can proceed. We will limit ourselves to the computation of the beta
function, which is our problem. The normalization conditions
(\ref{e:norm}) fix the zero--loop terms in the loop--wise expansion of
$g_i$, $m_0$ and $Z$:
\begin{eqnarray*}
g_i^R&&=g_i^{(0)}\nonumber \\
m^2&&=m^{2\, (0)}_0 \nonumber \\
Z^{(0)}&&=1.
\end{eqnarray*}
One--loop calculations are easily done. Of course $m_0^{2\,(1)}=0$ and
$Z^{(1)}=0$. This implies that up to one loop $\gamma_1 =0$ (that is
$\gamma_1=O(g^2)$). It is convenient to write down the results for the
couplings in terms of $g_4$, $g_{1\perp}$, $g_2$ and $\tilde
g\equiv g_2-g_{1\parallel}$. We find:
\begin{eqnarray}\label{e:uno}
g_{1,\perp}^{(1)}&&=-\frac{1}{\pi}\left[ \ln m -\left(\ln 2 +\ln p
-1\right)\right]g_{1\perp}^{R}g_{2}^R\nonumber \\
&&+\frac{1}{\pi}\left[ \ln m
-\left(\ln 2 +\ln p -\frac{1}{2}\right)\right]g_{1\perp}^{R}\tilde g^R
-\frac{1}{2\pi}g_{1\perp}^R g_{4}^R\nonumber \\
g_2^{(1)}&&=-\frac{1}{2\pi}\left[ \ln m -(\ln 2 +\ln p
-1)\right]g_{1\perp}^{R\,2} \nonumber \\ \tilde
g^{(1)}&&=\frac{1}{2\pi}\left[ \ln m -\ln \Lambda
+\frac{1}{2}(1+\mbox{\boldmath $C$} + \ln 2)\right] g_{1\perp}^{R\,2}
\nonumber \\ g_{4}^{(1)}&&=-\frac{1}{4\pi}g_{1\perp}^{R\,2},
\end{eqnarray}
where {\boldmath $C$} is the Euler constant. In Eq. (\ref{e:uno}) we
note the presence of $p$. However simply on the basis of dimensional
analysis we can exclude that $p$ will appear in the final
result. Two--loop calculations are tedious and we will omit the
details. We calculate only the singular terms in $m$ since we do not
plan to go beyond the two--loop approximation. We report the results
for $Z$ and $g_{1\perp}$:
\begin{equation}\label{e:Z}
Z^{(2)}=\frac{1}{2^3\pi^2}\ln m
\left(g_{1\perp}^{R\,2}+g_{1\parallel}^{R\,2}+2g_{2}^{R\,2}-2g_{1\parallel}^{R}g_{2}^{R}\right)
\end{equation}
\widetext
\begin{eqnarray}\label{e:due}
&& g_{1,\perp}^{(2)} = \frac{1}{2\pi^2}\left[ \ln^2 m -2\ln m
\left(\ln 2 +\ln p -1\right)\right]g_{1\perp}^{R}g_{2}^{R\,2} \nonumber
\\
&& +\left\{
\frac{1}{2\pi^2}\ln m -\frac{1}{\pi^2}\left[ \ln^2 m -2\ln m \left(\ln
2 +\ln p -\frac{3}{4}\right)\right]\right\}g_{1\perp}^{R}g_{2}^{R}\tilde
g^R\nonumber \\ &&+\frac{1}{2\pi^2}\ln
m\,g_{1\perp}^{R}g_{2}^{R} g_4^R-\frac{1}{2\pi^2}\ln
m\,g_{1\perp}^{R}\tilde g^{R} g_4^R+ \frac{1}{2\pi^2}\left[ \ln^2 m
-2\ln m \left(\ln 2 +\ln p -\frac{3}{4}\right)\right]g_{1\perp}^{R\,3}
\nonumber \\ &&+\frac{1}{2\pi^2}\left[ \ln^2 m -2\ln m \big(\ln 2
+\ln p -\frac{1}{2}\big)\right]g_{1\perp}^{R}\tilde g^{R\,2} -
\frac{1}{2^2\pi^2}\ln m
\left(g_{1\perp}^{R\,2}+g_{1\parallel}^{R\,2}+2g_{2}^{R\,2}-2g_{1\parallel}^{R}g_{2}^{R}\right)
\end{eqnarray}
\narrowtext From Eqs. (\ref{e:due}) and (\ref{e:uno}) we derive
$\beta^{(2)}_{1\perp}(g^R)$. The final result is
\begin{equation}\label{e:CS-ris}
\beta_{1\perp}(g^R) = \frac{1}{\pi}
g_{1\parallel}^Rg_{1\perp}^R +
\frac{1}{4\pi^2}(g_{1\parallel}^{R\,2}g_{1\perp}^R
+g_{1\perp}^{R\,3}) +O(g^4).
\end{equation}
It can be noted that $p$ does not appear in Eq. (\ref{e:CS-ris}), as
expected.  A crucial use in deriving Eq. (\ref{e:CS-ris}) is made of
Eq. (\ref{e:Z}), which is responsible for the cancellations expected
from the exact solutions of the Luttinger and Mattis models
\cite{lut,ml,mat3}, and for the anomalous behavior of the theory. Of
course the anomalous exponent $\eta=\gamma_1(g^\star)$ derived from
Eq. (\ref{e:Z}) when $0<g_1 \ll 1$ is in agreement with the exact
solution of the Luttinger model, where $g_1=0$, $g_4=0$.

Equation (\ref{e:CS-ris}) is the same as Eq. (\ref{e:lie}) for the
$g_{1\perp}$ coupling. For $g_2$ and $g_{1\parallel}$ the same
conclusion holds: the beta function of the GML method is
recovered. The present CS approach, admittedly too involved, has
perhaps the only value in that no use is made of approximate
multiplicative relations.

\section{The Wilson approach}
The multiscale formulation of the Wilson RG\cite{gal} is particularly
well suited to study the running of the coupling constants by discrete
steps. The application of this method to interacting one--dimensional
fermionic systems started with Refs.\onlinecite{bg,bgps} and was thoroughly
developed and applied to various problems\cite{bmg,rc1,rc2,rc3}.
Here we give a short and simplified account of the
method and refer to the cited papers for the details.

In the coordinate space the free propagator $G_\omega(x)$ for $\omega$
particles (again $\omega=\pm1$ and $v_F=1$) is:
$$
G_\omega(x)=\frac{1}{(2\pi)^2}\int dk_0\,dk_1
\frac{e^{-i(k_0x_0+k_1x_1)}}{-ik_0 + \omega k_1}.
$$
Actually it is not necessary to start with a kinetic term linearized
around the Fermi surface: the RG can deal with realistic quadratic
dispersion relations \cite{bg}. This simplification is however inessential
for our purposes. Let $p$ be an arbitrary momentum scale which for instance
may be chosen equal to the inverse of the range of the potential.
The propagator is decomposed in the sum 
$$
G_\omega(x) = \sum_{h=-\infty}^{1}G^{(h)}_{\omega}(x),
$$
with 
\begin{eqnarray}
G_\omega^{(1)}(x)&&=\frac{1}{(2\pi)^2}\int
dk\,e^{-ikx}\frac{1-e^{-p^{-2}(k_0^2+k_1^2)}}{-ik_0 + \omega
k_1},\nonumber \\
G^{(h)}_\omega(x) &&= \frac{1}{(2\pi)^2}\int dk\, \frac{e^{-ikx}}{-ik_0 +
\omega k_1}\nonumber \\
&& \times\left[ e^{-p^{-2}\gamma^{-2h}(k_0^2 + k_1^2)} -
e^{-p^{-2}\gamma^{-2h+2}(k_0^2 + k_1^2)} \right], \label{e:prop}
\end{eqnarray}
where $h\le0$, $\gamma >1$ and $kx=k_0x_0 + k_1x_1$.  This decomposition
divides the u.v. from the i.r. singularity of the propagator:
$G_\omega^{(1)}$ is singular in the u.v.  while
$\sum_{h=-\infty}^0G_{\omega}^{(h)} = G_{\omega}^{i.r.}$ is singular
in the i.r. It is important to note that $G^{(h)}_\omega(x)$, the
propagator on scale $h$, for $h\le 0$ has an u.v. and an i.r. cutoff:
$G^{(h)}_\omega(x)$ is essentially different from $0$ only for $x\sim
\gamma^{-h}$ ($k\sim \gamma^h$ in momentum space).

One imagines that this decomposition stems from a similar
decomposition of the fields:
$$
\psi^{\pm}_{x,\omega,\sigma}=\sum_{h=-\infty}^{1}\psi^{\pm\,(h)}_{x,\omega,\sigma}
$$
such that the pairings in the Grassmannian Wick rule are
\begin{eqnarray*}
\int P(d\psi^{(h)}_\omega)\,
&&\psi^{+\,(h)}_{x,\omega,\sigma}\psi^{-\,(h')}_{y,\omega',\sigma'}\equiv
\langle \psi^{+\,(h)}_{x,\omega,\sigma}\psi^{-\,(h')}_{y,\omega',\sigma'}
\rangle \\ && \qquad \equiv
\delta_{\omega,\omega'}\delta_{\sigma,\sigma'}\delta_{h,h'}G^h_\omega(x-y).
\end{eqnarray*}
We are interested to study the i.r. effective potential $V^{(0)}$
arising from the integration of the u.v. component $\psi^{(1)}_\omega$
from the effective potential $V_{\text{eff}}(\varphi)$ defined by:
$$
e^{-V_{\text{eff}}(\varphi)} = \frac{1}{\mathcal{N}}\int P(d\psi)\,
e^{-V(\psi+\varphi)},
$$
where $\mathcal N$ is a normalization constant and $V$ is the
interaction potential. The ultraviolet integration was actually
performed for the spinless model\cite{bgps}. In the following we
suppose to start directly with $V^{(0)}$.

The core of the method consists of a procedure that, integrating out
the fields from the higher to the lower \textit{scales} $h$ ($h\to -
\infty$), constructs a well defined dynamical system of running
coupling constants $g_h$, whose iteration map is the beta
\textit{functional}.

The operators $\mathcal L$ and ${\mathcal R}=1-{\mathcal L}$ are
introduced. ${\mathcal R}$ is the usual renormalization operator of
the BPHZ scheme: its action on a given vertex $\Gamma$, in momentum
space for instance, is given by ${\mathcal R}(\Gamma)=\Gamma
-t^\Gamma(\Gamma)$, where $t^\Gamma$ denotes the Taylor series with
respect to the external momenta of $\Gamma$ up to order $D(\Gamma)$,
if $D(\Gamma)$ is the $\Gamma$ superficial degree of divergence. Of
course ${\mathcal L}(\Gamma)=t^\Gamma(\Gamma)$.

The couplings $g_h$ on a given scale $h$ are defined by an inductive
scheme. Let us assume we have constructed the effective potential
$V^{(h)}(\psi_\omega^{(\le h)}, g_{h+1},\ldots,g_0)$ on scale $h$,
where $\psi_\omega^{(\le h)}=\sum_{n\le h}\psi^{(n)}_\omega$ and
$g_{h+1},\ldots,g_0$ are the previously defined couplings on scales
$h+1,\ldots,0$, . We define
$$
\overline{V}^{(h)}(\psi_\omega^{(\le h)},g_h)\equiv {\mathcal L}
V^{(h)}(\psi_\omega^{(\le h)}, g_{h+1},\ldots,g_0).
$$
The previous relation introduces the $g_h$ and relates them to the
$g_{h+1},\ldots,g_0$ through the beta functional $B_h$: $g_h =
g_{h+1}+B_h(g_{h+1},\ldots,g_0)$. The effective potential $V^{(h-1)}$
on scale $h-1$ is defined by
\begin{eqnarray}\label{e:Vh-1}
&&e^{-V^{(h-1)}(\psi^{(\le h-1)})} \equiv \nonumber \\
&&\qquad \frac{1}{{\mathcal N}'}\int
P(d\psi^{(h )})\, e^{-{\mathcal L}V^{(h)}(\psi^{(\le h)}) -{\mathcal
R}V^{(h)}(\psi^{(\le h)})}.
\end{eqnarray}
Of course $V^{(h-1)}=V^{(h-1)}(\psi_\omega^{(\le h-1)},
g_{h},\ldots,g_0)$. The procedure is then iterated. The starting point
is given by the couplings $g_0$ of ${\mathcal L} V^{(0)}$.  The final goal
is to find a region in the space of parameters $g_0$ where each initial
value generates a trajectory $g_h = g_{h+1}+B_h(g_{h+1},\ldots,g_0)$
such that the Schwinger functions are analytic in the $g_h$.

Unfortunately this scheme in our problem requires emendation. From the
second order result it becomes clear that $\alpha_h$ and $\zeta_h$ grow
too fast independently on the initial conditions. The point is that we
know that the interacting propagator has an anomalous behavior:
asymptotically for large distances it decays faster than the free
propagator. The wavefunction renormalization necessary to cure this
problem is accomplished by an inductive procedure that redefines step
by step the free measure of the functional integral and the couplings
by the means of a sequence of parameters $Z_{h}$ with
$h=0,-1,\ldots$. Let us assume we have introduced
$Z_h,Z_{h+1},\ldots,Z_0$ and applied our procedure integrating out the
scales from $0$ to $h+1$ ($h < 0$). We get an effective potential
$\tilde V^{(\le h)}$ (different from $V^{(\le h)}$, defined by
Eq. (\ref{e:Vh-1})). We denote with $P_{Z_h}(\psi_\omega^{( h)})$,
$P_{Z_h}(\psi_\omega^{( \le h-1)})$ and $\tilde
P_{Z_h}(\psi_\omega^{(h)})$ the free measures with propagators,
respectively, $G_\omega^{(h)}/Z_h$, $G_\omega^{(\le h-1)}/Z_h$ and
$\tilde G_\omega^{(h)}/Z_h$, where the last one is the modified
propagator on scale $h$ and $G_\omega^{(\le h-1)}=\sum_{i\le
h-1}G_\omega^{(i)}$. $\hat V^{(h-1)}$ is defined by
\begin{eqnarray}\label{e:inizio}
&&\int P_{Z_{h}}(d\psi_\omega^{(\le h -1)})\,e^{-\hat
V^{(h-1)}(\sqrt{Z_{h}}\psi_\omega^{(\le h-1)})} =\nonumber \\
&& \quad \int
P_{Z_{h}}(d\psi_\omega^{(\le h
-1)})\tilde P_{Z_h}(d\psi_\omega^{(h)})\,e^{-\tilde V^{(\le
h)}(\sqrt{Z_{h}}\psi_\omega^{(\le h)})}.
\end{eqnarray}
$\hat V^{(h-1)}(Z_{h}^{1/2}\psi^{(\le h-1)})$ has the form:
\begin{eqnarray*}
&& \hat V^{(\le h-1)}(Z_{h}^{1/2}\psi_\omega^{(\le h-1)}) = ({\mathcal
L} + {\mathcal R})\hat V^{(\le h-1)}(Z_{h}^{1/2}\psi_\omega^{(\le
h-1)})\\ && = Z_{h}\Bigg\{\nu_{h-1}\sum_{\omega,\sigma} \int
\frac{d^2k}{(2\pi)^2}\,\psi^{(\le h-1)\,+}_{k,\omega,\sigma}\psi^{(\le
h-1)\,-}_{k,\omega,\sigma}\\ && +\zeta_{h-1}\sum_{\omega, \sigma} \int
\frac{d^2k}{(2\pi)^2}\,\psi^{(\le h-1)\,+}_{k,\omega,\sigma}(-i
k_0)\,\psi^{(\le h-1)\,-}_{k,\omega,\sigma}\nonumber \\
&&+\alpha_{h-1}\sum_{\omega, \sigma} \int
\frac{d^2k}{(2\pi)^2}\,\psi^{(\le h-1)\,+}_{k,\omega,\sigma}(\omega
k_1)\,\psi^{(\le h-1)\,-}_{k,\omega,\sigma}\Bigg\} + \ldots
\end{eqnarray*}
Now we add and subtract from $\hat V^{(h-1)}$ the term $\propto
Z_h\zeta_{h-1} \psi^{+ (\le h-1)}_{k,\omega,\sigma} (\omega
k_1)\,\psi^{- (\le h-1)}_{k,\omega,\sigma}$ and insert the term
$\propto Z_h \zeta_{h-1} \psi^{+ (\le h-1)}_{k,\omega,\sigma}
(-ik_0+\omega k_1)\,\psi^{- (\le h-1)}_{k,\omega,\sigma}$ in the free
measure. Let $P'_{Z_h}(\psi_\omega^{( \le h-1)})$ be the measure
changed this way. We define $Z_{h-1}=Z_{h}(1+\zeta_{h-1})$ and write:
\widetext
\begin{eqnarray}
&&\int P_{Z_h}(d\psi_\omega^{(\le
 h-1)})\,e^{-\hat V^{(h-1)}(\sqrt{Z_{h}}\psi_\omega^{(\le
h-1)})}= \int P'_{Z_h}(d\psi_\omega^{(\le
 h-1)})\,e^{-V'^{(h-1)}(\sqrt{Z_{h}}\psi_\omega^{(\le h-1)})}\label{e:fine1} \\
=&&\int P_{Z_{h-1}}(d\psi_\omega^{(\le
 h-2)})\tilde P_{Z_{h-1}}(d\psi_\omega^{(
 h-1)})\,e^{-V'^{(h-1)}(\sqrt{Z_{h}}\psi_\omega^{(\le
h-1)})}\label{e:fine2} \\
=&&\int P_{Z_{h-1}}(d\psi_\omega^{(\le
h -2)})\tilde P_{Z_{h-1}}(d\psi_\omega^{(
h-1)})\,e^{-\tilde V^{(h-1)}(\sqrt{Z_{h-1}}\psi_\omega^{(\le
h-1)})}\label{e:fine3}.
\end{eqnarray}
\narrowtext
In Eq. (\ref{e:fine1}) $V'^{(h-1)}$ is obtained from $\hat V^{(h-1)}$
dropping the $\zeta_{h-1}$ term and substituting
$\alpha_{h-1}-\zeta_{h-1}$ for $\alpha_{h-1}$. In Eq. (\ref{e:fine2}) $\tilde
P_{Z_{h-1}}(d\psi_\omega^{( h-1)})$ is the free measure with propagator
$\tilde G_\omega^{(h-1)}/{Z_{h-1}}$ defined such that the remaining
part of the free measure is exactly $P_{Z_{h-1}}(d\psi_\omega^{(\le h
-2)})$. Finally Eq. (\ref{e:fine3}) defines $\tilde V^{(h-1)}$ and the
r.h.s has the same structure of the r.h.s. of Eq. (\ref{e:inizio}) so
the procedure may be reiterated. The starting point are $Z_0=1$ and
$\tilde G_\omega^{(0)}=G_\omega^{(0)}$. 
The relations between the couplings of ${\mathcal L}\hat V^{(\le
h-1)}(\psi_\omega^{(\le h-1)})$, $g'_{i,h-1}$ ($i=1,2,4$),
$\alpha_{h-1}$ and $\zeta_{h-1}$, and the couplings of ${\mathcal
L}\tilde V^{(\le h-1)}(\psi_\omega^{(\le h-1)})$, $g_{i,h-1}$ and
$\delta_{h-1}$, are easily found:
\begin{eqnarray*}
&&\delta_{h-1} = \frac{Z_{h}}{Z_{h-1}}(a_{h-1}-z_{h-1}) \\
&&g_{i,h-1}=\left(\frac{Z_{h}}{Z_{h-1}}\right)^2g'_{i,h-1} \quad i=1,2,4.
\end{eqnarray*}
Of course $\zeta_{h-1}$ is no more present in $\tilde V^{(\le
h-1)}(\psi_\omega^{(\le h-1)})$, but $Z_{h-1}$ is introduced. The
replacement $\alpha_h \to \delta_h$ drastically improves the
convergence properties in the limit $h \to -\infty$.

Now we have the full recipe to proceed. Needless to say a crucial use
of the linked cluster theorem will be made.  For brevity only the
calculation for the $g_1$ coupling is sketched (we take
$g_{1\perp}=g_{1\parallel}=g_1$).

Let $C_{i,j}$ denote the loop of figure \ref{fig2}:
\begin{equation}\label{e:notazioni}
C_{i,j}=C_{i-j}=\int
\frac{d^2k}{(2\pi)^2}\,G^{(i)}_\omega(k)G^{(j)}_{-\omega}(k),
\end{equation}
where $i,j\le 0$ and $G_\omega^{(i)}$ is the propagator in momentum
space on scale $i$. It is immediate to verify that the r.h.s. of
(\ref{e:notazioni}) does not depend on $\omega$ nor on the scale $p$
(see (\ref{e:prop})) and that it is a function only of the difference
$i-j$. In particular $C_{i,i}=C_0$, $i \le 0$. The second order
calculation gives for $g_1$:
\begin{equation}\label{e:2ord_1}
g_{1,h-1} = g_{1,h} + \frac{1}{2}K\bigg(C_{h,h}g_{1,h}^2 +2\sum_{h<j\le
0} C_{h,j}\,g_{1,j}^2\bigg).
\end{equation}
$K$ is a combinatorial factor: $K=4$.  The previous relations give the
second order approximation for the beta \textit{functional}. Since we
are interested in the beta \textit{function}, in Eq. (\ref{e:2ord_1})
we write the $g_{1,j}$ for $j>h$ as functions of $g_{1,h}$. This
inversion generates a correction to the higher orders, in particular
to the third order.  We have (up to $O(g_1^4)$ terms):
\begin{eqnarray}\label{e:2ord_corr}
&& g_{1,h-1} = g_{1,h} +\frac{1}{2}K\bigg(C_{h,h}
+2\sum_{h<j\le 0}\! C_{h,j}\bigg)g_{1,h}^2\nonumber \\
&&-K^2\sum_{h<i\le 0}C_{h,i}\bigg(\sum_{h<j\le i}\!C_{j,j}
+2\!\sum_{h<k\le i}\sum_{k<j\le 0} C_{k,j}\bigg)g_{1,h}^3.
\end{eqnarray}
Using Eq. (\ref{e:prop}) for $G_\omega^{(h)}$ we find:
\begin{eqnarray}\label{e:2ord_2}
&& C_{h,h} +2\sum_{h<j\le 0} C_{h,j} = \frac{1}{2\pi}[ +\ln \gamma -
\ln \left( 1+\gamma^{2-2h}\right)\nonumber \\
&& + \ln \left(
1+\gamma^{-2h}\right)] = -\frac{1}{2\pi} \ln \gamma +
O(\gamma^{2h}).
\end{eqnarray}
From Eqs. (\ref{e:2ord_corr}) and (\ref{e:2ord_2}) we get the second
order \textit{discrete} beta function, up to $\gamma^{2h}$ terms (let's
remember that $\gamma >1$ and that we are interested in the $h\to
-\infty$ limit):
$$
g_{1,h-1} = g_{1,h} -\frac{\ln \gamma}{\pi} g_{1,h}^2 + O(g^3).
$$
The second line of Eq. (\ref{e:2ord_corr}) gives the corrections to the
third order result. We find:
\begin{eqnarray*}
&&\sum_{h<i\le 0}C_{h,i}\bigg(\sum_{h<j\le i}C_{j,j}
+2\sum_{h<k\le i}\sum_{k<j\le
0} C_{k,j}\bigg)=\\
&& \frac{1}{2\pi}\sum_{h<i\le 0}C_{h,i}\left[ -\ln
\left(1+\gamma^{2i}\right)+ (i-h)\ln \gamma \right], 
\end{eqnarray*}
where the r.h.s. is easily
calculated:
\begin{eqnarray}
&&\biggl\vert\sum_{h<i\le 0}C_{h,i}\ln
\left(1+\gamma^{2i}\right)\biggr\vert \le A\gamma^{h}\sum_{h<i\le0}1\, ,
\label{e:trucco}\\
&&\sum_{h<i\le 0}(i-h)C_{h,i}= \frac{\ln 2}{4\pi}\label{e:ln2}.
\end{eqnarray}
$A$ is a constant. We neglect the r.h.s of
Eq. (\ref{e:trucco}).  Equation (\ref{e:ln2}) too is derived
neglecting terms exponentially vanishing with $h$. Putting all
together we find a correction $\mathbf c$ to the third order given by:
$$
{\mathbf c} = -\frac{2\ln 2\ln \gamma}{\pi^2}g_{1,h}^3.
$$
The computation of the third order is tedious. We will limit ourselves
to note that exists a contribution $\propto \ln 2\ln \gamma$ of
two--loop diagrams that cancels exactly $\mathbf c$. This is important
because there is no such term in Eqs. (\ref{e:lie}), (\ref{e:gamma})
or (\ref{e:CS-ris}).  It comes from the diagrams $D_1$ and $D_2$ of
figure \ref{fig3}, which are related by $D_1=-2D_2$. We will consider
the simpler $D_1$. The contributions to $D_1$ given by
\begin{eqnarray*}
&& \sum_{h<i\le 0}C_{h,h}2C_{h,i} +\sum_{h<i\le 0}\sum_{h<j<
i}2C_{h,i}2C_{h,j}\\
&&+ \sum_{h<i\le 0}\sum_{h<j,k<1}C_{h,i}2C_{j,k}
+\sum_{h<i\le 0}\sum_{h<j<i}\sum_{h<k\le j}2C_{h,j}2C_{k,i}
\end{eqnarray*}
amount to 
\begin{equation}\label{e:fine}
\frac{1}{(2\pi)^2}\ln 2\ln\gamma
-\frac{1}{2\pi}\sum_{h<i<0}2C_{i,h}\ln\left(1+\gamma^{2i} \right).
\end{equation}
The second term of Eq. (\ref{e:fine}) can be neglected as for
Eq. (\ref{e:trucco}) and the first one gives the desired cancellation
(taking into account the combinatorial factors). The final result is
the same as Eq. (\ref{e:gamma}) or (\ref{e:CS-ris}) with
$g_{1_\parallel}=g_{1\perp}=g_1$. Again we find the fixed point
$g_1^\star$ of Eq. (\ref{e:fisso_g}) and we recover Eqs. (\ref{e:lie})
in the limit $\gamma\to 1$.

\section{Conclusions}
We have computed the third order (two--loop) approximation of the beta
function for a one dimensional model of interacting fermions, aiming
in particular to study the case of attractive interaction. An existing
result\cite{sol,sol2}, derived making use of tacitly assumed
approximations, pointed out a $O(1)$ fixed point. We tried to support
this conclusion setting a Callan--Symanzik approach and using the
Wilson RG formulated as in Refs.\onlinecite{gal,bg,bgps}. In each case we
recovered the aforementioned result.

An attempt to pursue further the examination of the problem was made
in Refs.\onlinecite{ting,q1,q2} where the fourth--order approximation,
which is by no means universal, was computed. We propose a different
approach focused on the study of the dependence on $\gamma$, the
rescaling factor of the RG group. A similar idea was discussed in
Ref.\onlinecite{wil}, where the dependence of the fixed points on the
parameters of the RG was analyzed. Our simple idea is that the
stability of the result with respect to $\gamma$ should indicate how
reliable one should consider the perturbative result.  It was expected
a third order result dependent on $\gamma$ but we found a too strong
dependence: the fixed point happens to change sign if $\gamma > \sqrt
e$, which is still $\sim 1$ (of course taking $\gamma \gg 1$ and, at
the same time, truncating at the third order would be
questionable). Nothing similar happens to the nontrivial fixed points
for repulsive interaction $\{g_1^\star=0,g_2^\star\}$, which are in
some sense insensitive to the value of $\gamma$.

Which conclusions can be drawn? It is useful to compare the one
dimensional interacting Fermi Gas with the well known Kondo
problem. It can be noted that the scaling equation for $g_1$ (Fermi
Gas) and the one for the impurity coupling $\lambda$ (Kondo model) have the
same structure (see e.g. Ref.\onlinecite{affl} for a review in the modern
language of Conformal Field Theory). The Kondo effect was thoroughly
investigated. The ferromagnetic case corresponds to the Fermi gas with
repulsive interaction ($g_1>0$): the RG flow is such that $\lambda \to
0$. If the coupling is antiferromagnetic the system flows toward a
strong coupling phase ($\lambda \to \infty$)\cite{wil2}. Moreover the
infrared divergences induce a scale, the Kondo temperature $T_{\text{K}}$,
characterizing the low energy physics.

In our case the particular instability of the perturbative result with
respect to $\gamma$, a dimensionless parameter without a physical
meaning, should indicate that the RG flow does not actually stop at a
finite value and suggests a conclusion similar to the previous one. In
our case the characteristic scale should be a gap $\Delta$ for the
spin degrees of freedom, whose expression, according to
Ref. \onlinecite{lar}, should have for small coupling an expression of
the type $\Delta \propto \sqrt{g_1}\exp(-1/g_1)$.

\acknowledgments This work received support from the INFN, section of
Cagliari, and from the ``Dipartimento di Scienze Fisiche'' of Cagliari
University. We are deeply indebted with Prof.  G. Gallavotti, from
whom we learned the multiscale formulation of the RG, for having
originated the theme of this research as well as for many
discussions. We acknowledge useful discussions with Prof. E. Marinari,
Dr. V. Mastropietro, Prof. G. Benfatto and Dr. M. Lissia.



\begin{figure}
\caption{One loop diagram contributing to the four--point vertex
function. The full (dashed) lines represent right (left) moving
fermions. The value of the graph is proportional to
$-\frac{1}{2\pi}\ln \left(\frac{k_0}{E_0}\right) +\frac{1}{4\pi}\ln
\left(1+\frac{k_0^2}{E_0^2} \right)$.}
\label{fig1}
\end{figure}

\begin{figure}
\caption{One--loop contribution defining $C_{i,j}$. $k$ is the
internal momentum. The two propagators are on scales $i$ and $j$. }
\label{fig2}
\end{figure}

\begin{figure}
\caption{Two--loop graphs contributing to the beta function: $D_1$
diagrams (a) and $D_2$ (b) give simply related contributes:
$D_1=-2D_2$.}
\label{fig3}
\end{figure}

\end{document}